\documentclass[a4paper,fleqn]{cas-sc}

\usepackage[authoryear,longnamesfirst]{natbib}

\usepackage{graphicx}
\usepackage{algorithm}
\usepackage{subcaption}
\usepackage{algpseudocode}
\usepackage{booktabs}
\usepackage[font=small,labelfont=bf]{caption}

\def\tsc#1{\csdef{#1}{\textsc{\lowercase{#1}}\xspace}}
\tsc{WGM}
\tsc{QE}
\tsc{EP}
\tsc{PMS}
\tsc{BEC}
\tsc{DE}

\begin{document}
\let\WriteBookmarks\relax
\def\floatpagepagefraction{1}
\def\textpagefraction{.001}

\shorttitle{Leveraging social media news}

\shortauthors{S. Liang et~al.}

\title [mode = title]{Real-Time Lane-Level Crash Detection on Freeways Using Sparse Telematics Data}                      










\author[1]{Shixiao Liang}
\ead{sliang85@wisc.edu}

\author[1]{Chengyuan Ma*}
\ead{cma97@wisc.edu}

\author[2]{Pei Li, Ph.D.}
\ead{pei.li@uwyo.edu}

\author[3,1]{Haotian Shi}
\ead{shihaotian95@tongji.edu.cn}

\author[1]{Jiaxi Liu}
\ead{jliu2487@wisc.edu}

\author[1]{Hang Zhou}
\ead{hzhou364@wisc.edu}

\author[1]{Keke Long}
\ead{klong23@wisc.edu}

\author[4]{Bofeng Cao}
\ead{hzhou364@wisc.edu}

\author[5]{Todd Szymkowski}
\ead{todd.szymkowski@dot.wi.gov}

\author[1]{Xiaopeng Li}
\ead{xli2485@wisc.edu}



\affiliation[1]{organization={Department of Civil \& Environmental Engineering, University of Wisconsin-Madison},
    city={Madison},
    state={WI},
    country={United States}}

\affiliation[2]{organization={Department of Civil \& Architectural Engineering and Construction Management, University of Wyoming},
    city={Laramie},
    state={WY},
    country={United States}}
    
\affiliation[3]{organization={Department of Traffic Engineering \& Key Laboratory of Road and Traffic Engineering of Ministry of Education, Tongji University},
    city={Shanghai},
    country={China}}
    
\affiliation[4]{organization={Department of Electrical \& Computer Engineering, University of Wisconsin–Madison},
    city={Madison},
    state={WI},
    country={United States}}

\affiliation[5]{organization={Wisconsin Department of Transportation},
    city={Madison},
    state={WI},
    country={United States}}


\begin{abstract}
Real‐time traffic crash detection is critical in intelligent transportation systems because traditional crash notifications often suffer delays and lack specific, lane‐level location information, which can lead to safety risks and economic losses. This paper proposes a real‐time, lane‐level crash detection approach for freeways that only leverages sparse telematics trajectory data. In the offline stage, the historical trajectories are discretized into spatial cells using vector cross‐product techniques, and then are used to estimate an vehicle intention distribution and select an alert threshold by maximizing the F1‐score based on official crash reports. In the online stage, incoming telematics records are mapped to these cells and scored for three modules: transition anomalies, speed deviations, and lateral maneuver risks, with scores accumulated into a cell‐specific risk map. When any cell’s risk exceeds the alert threshold, the system issues a prompt warning. Relying solely on telematics data, this real‐time and low‐cost solution is evaluated on a Wisconsin dataset and validated against official crash reports, achieving a 75\% crash identification rate with accurate lane-level localization, an overall accuracy of 96\%, an F1-score of 0.84, and a non-crash–to-crash misclassification rate of only 0.6\%, while also detecting 13\% of crashes more than 3 minutes before the recorded crash time.
\end{abstract}


\begin{highlights}
\item A real-time crash detection framework that uses large-scale telematics data to achieve lane-level localization and reliable detection under sparse and low-frequency sampling.

\item A general and adaptive modeling mechanism that discretizes vehicle trajectories into lane-aligned cells and characterizes normal driving behaviors for robust lane-level crash detection.

\item Using large-scale Wisconsin telematics data and ground-truth crash reports, the detection algorithm achieved a 75\% crash identification rate, with only 0.6\% of non-crashes misclassified as crashes, while also detecting 13\% of crashes more than 3 minutes before the record time in crash report.
\end{highlights}


\maketitle

\section{Introduction}
Real-time traffic crash detection plays a critical role in enhancing traffic safety and enabling timely emergency responses \citep{huang2020highway,hussain2024integrating,yuan2018real, li2022hybrid,li2020application}. Reducing crash detection and clearance time by just one minute can yield significant safety and efficiency benefits. According to the \citet{FHWA2010Outreach}, each additional minute of crash duration increases the likelihood of secondary crashes by approximately 2.8\% , while reducing clearance time by one minute can cut total traffic delay by up to four minutes. These time savings also carry measurable economic significance: in Hong Kong, crash-related traffic delays in 2021 cost society approximately US \$11.02 million, corresponding to an average of about US \$0.26 per vehicle-minute of lost productivity and fuel consumption \citep{lian2024cost}. Thus, even a one-minute improvement of crash detection and response could prevent multiple secondary crashes and save substantial societal costs, underscoring the critical importance of rapid, accurate crash alerts.

Road traffic crash detection refers to the task of identifying when and where incidents occur based on available traffic data (e.g., aggregated traffic measures or vehicle trajectories). Current crash detection approaches fall into two broad categories based on the criteria of detection. \citep{elsahly2022systematic}. The first category focuses on using macroscopic, aggregated traffic indicators such as flow, density, and speed, to identify abnormal traffic conditions \citep{zhu2021dynamic,zhang2023accident,querfurth2025crash}. While effective at identifying major crashes that cause network-wide congestion, these techniques often suffer from substantial detection delays because they rely on traffic disturbances gradually propagating through the network. Their spatial resolution is inherently coarse, typically limited to corridor-level anomalies rather than precise lane-level roadway segments, which highlights a persistent gap in fine-grained crash localization research. Existing macroscopic studies rarely evaluate the exact spatial accuracy of crash localization, leaving the problem underexplored. 
Nonetheless, many operational and safety applications require much finer 
spatial resolution than macroscopic indicators can provide, because effective 
countermeasures rely on knowing the exact lane affected by an incident. 
Achieving lane-level precision is therefore critical for enabling proactive 
lane changes, maintaining smoother traffic flow, and reducing the likelihood 
of secondary crashes.

On the other hand, microscopic methods employ either roadside sensing or vehicle trajectory data to detect abnormal maneuvers at the lane level \citep{rocky2024review}. Approaches relying on roadside cameras, LiDAR, or other fixed sensors can provide highly accurate localization but require costly infrastructure and dense deployments, which limits scalability \citep{qu2024expressway}. Methods based on high-fidelity vehicle trajectories achieve comparable precision but often depend on proprietary or high-frequency data that are difficult to obtain in real time. Moreover, even when high-quality vehicle trajectory data are available, few studies have achieved lane-level accuracy without relying on additional sensor fusion. Therefore, these limitations highlight the need for a scalable and cost-effective data source that can capture vehicle-level dynamics while avoiding the deployment burden of dense roadside sensors or proprietary high-frequency datasets.

Telematics data, collected from in-vehicle GPS and communication systems, can provide time-stamped trajectories and speed profiles of individual vehicles, enabling microscopic observation of traffic dynamics at a large scale \citep{herfort2023spatio,kan2019traffic}. Recent industry reports further indicate that telematics systems have already reached an estimated 79 percent penetration among new vehicles worldwide in 2024, suggesting that such data sources are becoming increasingly ubiquitous \citep{berginsight2025oem}. 
Owing to their wide availability and low deployment cost, such data support 
low-latency, real-time crash detection at lane-level granularity while 
remaining scalable to nationwide deployment. However, although modern GPS technology can achieve sub-meter positioning accuracy, the trajectories collected through large-scale telematics platforms are typically temporally sparse because most providers transmit data at intervals of three to five seconds. On freeways, where vehicles travel at high speeds, this upload frequency results in long spatial gaps between consecutive points. In addition, only about five to seven percent of vehicles in the total traffic stream are represented in these data, which further limits the completeness of the reconstructed traffic state \citep{berginsight2025oem}. Leveraging such incomplete trajectories to identify abnormal traffic patterns remains challenging. To improve spatial precision, scholars have begun developing machine learning frameworks that snap sparse vehicle telematics records to detailed lane geometries for microscopic crash risk inference \citep{li2022real, kubin2021deep,dai2023human,qu2024towards, santos2022literature, huang2020highway,li2020real, li2022hybrid}. While these learning-based approaches achieve promising results, they often lack physical interpretability. Recent studies have explored physics-enhanced learning frameworks that combine data-driven models with physical priors \citep{lee2025physics, liang2025theory}, yet their implementation in crash detection still requires high-frequency data or specialized sensors (e.g., built-in accelerometers \citep{sharma2016s}). Despite these advances, a gap remains for lightweight and scalable methods that can operate effectively on sparse telematics data. Furthermore, most machine learning approaches rely on large labeled datasets and numerous contextual features (e.g., speed limits, weather, or infrastructure conditions), which constrains their generalizability and precision when applied solely to telematics data. 

Consequently, detecting crashes from sparse and low-granularity telematics trajectories therefore remains an open challenge\citep{pang2013detection, roy2024advance,kandiboina2025real}. To make effective use of these trajectories, one must extract and accumulate subtle abnormal driving behaviors such as sudden lane changes and decelerations that are often observed upstream of the crash site as drivers maneuver to bypass the blockage \citep{shi2025leveraging}. To address this, we propose a novel model-free framework that separates the offline model-building stage from the real-time detection stage to enable an efficient and adaptive crash detection process. In the offline stage, raw telematics records are matched to lane geometries and discretized into spatial cells to mitigate sampling sparsity. Historical trajectories are used to construct a vehicle intention distribution under normal conditions. The alert threshold is then determined by maximizing the F1-score, leveraging crash times, report times, and other relevant information from the official crash reports \citep{powers2020evaluation}. After preparing all these data, in the online detection stage, each incoming telematics record is assigned to its corresponding cell based on the same algorithm in the offline stage. For each assigned cell, we compute three risk metrics. The first metric measures how unexpected the observed transition is relative to the vehicle intention distribution. The second metric quantifies the degree of speed deviation from normal traffic conditions. The third identifies abnormal lane change maneuvers. These risk metrics are weighted summed to update a dynamic risk map every second. Whenever the accumulated risk in any cell exceeds the safety alarm threshold, the system issues a prompt warning with precise lane-level localization.

The main contributions of this study are as follows:

\begin{enumerate}
    \item A real-time crash detection framework that integrates large-scale telematics data for lane-level localization, enabling efficient crash detection even under sparse and low-frequency trajectories.  
    \item A general and adaptive modeling mechanism that spatially discretizes vehicle trajectories into lane-aligned cells and characterizes normal driving behaviors, providing a robust foundation for lane-level crash detection.  
    \item Using large-scale Wisconsin telematics data and ground-truth crash reports, the detection algorithm achieved a 75\% crash identification rate, with only 0.6\% of non-crashes misclassified as crashes, while also detecting 13\% of crashes more than 3 minutes before the record time in crash report.
\end{enumerate}

The remainder of this paper is organized as follows. Section~\ref{SEC_METHODOLOGY} details our methodology.  
Section~\ref{SEC_EXPERIMENT} describes the dataset and experimental setup and presents the results.
Section~\ref{SEC_CONCLUSION} concludes and outlines future research directions.

\section{Methodology}
\label{SEC_METHODOLOGY}
In this section, we address the problem of detecting freeway crashes in real time using sparse telematics data collected from CVs. The task relies on two types of information: (i) large-scale historical CV trajectories and official crash reports, which are used to learn normal driving patterns and to calibrate model parameters, and (ii) streaming real-time telematics data, which provide instantaneous observations of vehicle behavior. Our objective is to design a lane-level crash detection method to continuously monitor freeway segments and identify abnormal driving behaviors indicative of a crash. The final output of the methodology is a real-time crash detection signal that specifies whether a crash has occurred and, if so, the corresponding lane-level location.

\subsection{Framework Overview}
\begin{figure}
    \centering
\includegraphics[width=1\linewidth]{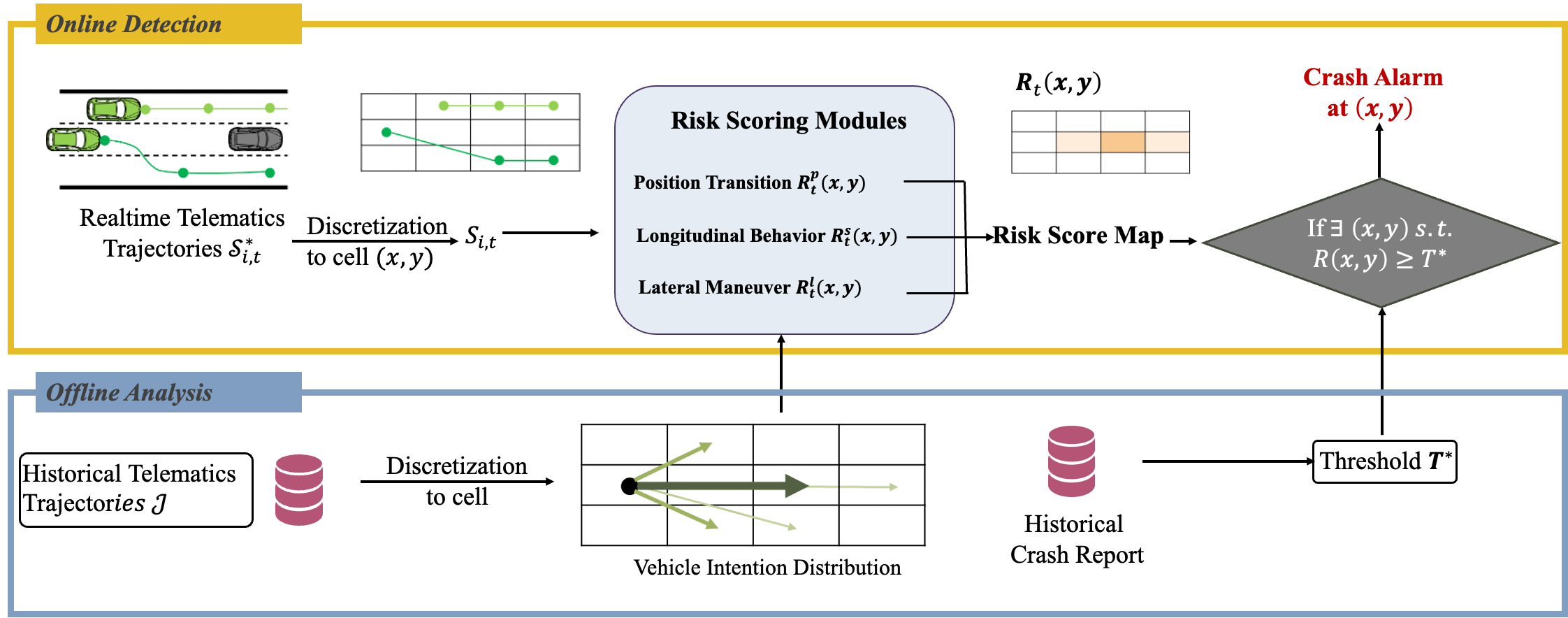}
\caption{Overview of the real‐time crash detection framework}
\label{fig:framework}
\end{figure}

Figure~\ref{fig:framework} illustrates the overall structure of the proposed crash detection framework, which consists of an offline analysis stage and an online detection stage. The offline stage relies on a large-scale historical telematics dataset $\mathcal{J}=\{S^*_{i,t}\mid i\in\mathcal{I},\, t=1,\dots,T\}$, where each raw telematics record is given by $S^*_{i,t}=(\mathrm{lat}_{i,t},\mathrm{lon}_{i,t},v_{i,t})$, denoting the reported latitude, longitude, and instantaneous speed of vehicle $i$ at time $t$. These records are transformed into discretized states 
$S_{i,t}=(x_{i,t},y_{i,t},v_{i,t})$, 
where $x_{i,t}$ denotes the lane index and $y_{i,t}$ denotes the longitudinal segment index on the roadway based on the map geometry of the study corridor. 
The set of all possible cell locations is denoted by 
$\mathcal{S} = \{(x,y)\}$, 
so that the spatial component of each discretized state satisfies $(x_{i,t},y_{i,t}) \in \mathcal{S}$.
 Using the historical dataset $\mathcal{J}$, the offline stage learns the vehicle intention distribution that characterizes normal longitudinal and lateral driving behaviors under non-crash conditions. In addition, official crash reports aligned with telematics trajectories are used to calibrate the alert threshold $T^*$ and other model parameters required for real-time detection.

During online operation, streaming telematics records $S^*_{i,t}$ arrive sequentially from a time-varying set of vehicles $i\in\mathcal{I}_t$. Each record is converted into a discretized state $S_{i,t}=(x_{i,t},y_{i,t},v_{i,t})$. For each time step, the system evaluates three instantaneous risk components at the visited cell $(x_{i,t},y_{i,t})$: the position transition risk $R^p_{i,t}$, the speed deviation risk $R^s_{i,t}$, and the lateral maneuver risk $R^\ell_{i,t}$. These per-vehicle risks are aggregated over all vehicles in the same cell and accumulated through time to form a dynamic risk map $R_t(x,y)$. Whenever a vehicle traverses a cell, the accumulated risk at that cell is reset, indicating that the region has been directly observed and verified to be safe. A crash alarm is issued when the accumulated risk at any cell exceeds the offline-calibrated threshold $T^*$, and the corresponding cell $(x,y)$ is reported as the detected crash location. Through this integration of historical intention patterns and real-time behavioral evidence, the framework achieves lane-level crash detection under sparse connected-vehicle penetration.

\subsection{Offline Calibration Stage}
\label{sec:offline}
\subsubsection{Discretization}

Discretization maps each raw telematics record 
$S^*_{i,t}=(\mathrm{lat}_{i,t},\mathrm{lon}_{i,t},v_{i,t})$
into a unique cell-level state 
$S_{i,t}=(x_{i,t},y_{i,t},v_{i,t})$, 
where $x_{i,t}$ is the lane index and $y_{i,t}$ is the longitudinal segment index.  
This mapping replaces free-floating GPS coordinates with a structured grid, reduces raw data complexity, supports efficient computation of transition probabilities, and provides the computational foundation for rapid online risk-score updates.  
Operating on a fixed grid further enables accumulated risk values to be visualized and updated with minimal overhead even under sparse CV penetration rate.

To determine the lane index $x_{i,t}$, we employ a lane-matching algorithm that assigns each telematics record to its corresponding lane using geometric rules based on publicly available map data such as OpenStreetMap (OSM).  
Traditional lane-detection techniques such as hidden Markov models, least-squares optimization, or fuzzy logic 
\citep{yang2018automatic, hansson2020lane} 
are computationally expensive, and image-based approaches  
\citep{guo2024lane, rabe2016lane}  
require continuous video streams unavailable in sparse CV data.  
Given improving GPS accuracy, a geometric method based on vector cross-product provides an efficient and robust alternative.

To perform lane matching, we first construct the reference geometry required for mapping each telematics record to its corresponding lane. Starting from the map-based centerline of the roadway, we obtain them from publicly available digital maps such as OSM. Based on these centerline points, we generate the reference line used for lane determination. For divided or one-way facilities, the centerline directly represents the travel direction; for bidirectional roadways, the centerline is laterally offset by half of the total carriageway width to create separate reference lines for each travel direction. Each reference line is then interpolated into a dense sequence of vertices to ensure consistent spatial resolution for projecting telematics points and computing lateral offsets.

With the reference geometry defined, each telematics record can be assigned to its lane using a geometric lane-matching procedure. For each record $S^*_{i,t} = (\mathrm{lat}_{i,t},\mathrm{lon}_{i,t},v_{i,t})$, we represent its position $(\mathrm{lat}_{i,t},\mathrm{lon}_{i,t})$ in the local roadway frame as point $A$. A perpendicular is dropped from $A$ to the selected reference line to obtain the foot point $C$. We then take the next telematics sample of the same vehicle in time$S^*_{i,t+1}$, denoting $(\mathrm{lat}_{i,t+1},\mathrm{lon}_{i,t+1})$ as point $B$, and form the vectors $\overrightarrow{CA}$ and $\overrightarrow{AB}$. Their cross product determines the direction of $A$ relative to the reference line: the sign of its $z$-component, following the right-hand rule, indicates whether $A$ lies to the left or right of the reference line.

The signed lateral offset is then computed as
$$\delta = \mathrm{signedDist}(A,C),$$
representing the perpendicular distance from $A$ to the reference line, with negative values on the left and positive values on the right. This continuous offset is discretized into the lane index $x_{i,t}$ using
\[
x_{i,t}
=
\left\lfloor\frac{\delta + \tfrac{W^*}{2}}{w}\right\rfloor
+ 1,
\]
where $W^*$ is the carriageway width associated with the selected reference line and $w$ is the nominal lane width. Adding $W^*/2$ shifts $\delta$ from $[-W^*/2,\,W^*/2]$ to $[0,\,W^*]$, and dividing by $w$ partitions the cross section into uniformly spaced lanes, with the floor operator producing a one-based integer lane index. The complete procedure is summarized in Algorithm~\ref{alg:lane_matching}.

\begin{algorithm}[H]
\caption{Lane Matching}
\label{alg:lane_matching}
\begin{algorithmic}[1]
\Require road segment of interest $(lat_c,lon_c)$, road identifier $ID_c$, telematics records $\mathcal{S}^*$, lane width $w$, lane count $n$, search radius $r$, OSM tags: \texttt{oneway}, \texttt{dual\_carriageway}
\State filter $\mathcal{S}'$ by proximity to $(lat_c,lon_c)$ within radius $r$
\State retrieve from OSM the centerline $C$ and tags \texttt{oneway}, \texttt{dual\_carriageway}
\State compute total width $W \gets w \times n$
\If{oneway = yes or dual\_carriageway = yes}
  \State $C^* \gets C,\quad W^* \gets W,\quad n^* \gets n$
\Else
  \State $d \gets W/4$
  \State $C_L \gets \text{offset}(C,+d),\quad C_R \gets \text{offset}(C,-d)$
  \State $C^* \gets \arg\min_{X\in\{C_L,C_R\}}\mathrm{dist}(X,(lat_c,lon_c))$
  \State $W^* \gets W/2,\quad n^* \gets n/2$
\EndIf
\ForAll{$A=S^*_{i,t} \in \mathcal{S}'$}
  \State $C_A \gets \arg\min_{c\in C^*}\mathrm{dist}(A,c)$
  \State $B \gets \text{next}(S^*_{i,t})$ \Comment{next telematics record of same vehicle $i$}
  \State $\vec{CA} \gets A - C_A$
  \State $\vec{AB} \gets B - A$
  \State $z \gets (\vec{CA} \times \vec{AB})_z$
  \If{$z>0$} $s\gets \mathrm{right}$ \Else $s\gets \mathrm{left}$ \EndIf
  \State $\delta \gets \mathrm{signedDist}(A,C_A)$
  \State $\mathrm{laneIdx} \gets \left\lfloor(\delta+W^*/2)/w\right\rfloor + 1$
  \State record $(\mathrm{laneIdx},s)$ for $A$
\EndFor
\State \textbf{Output:} laneIdx and side $s$ for all telematics records in $\mathcal{S}'$
\end{algorithmic}
\end{algorithm}

After assigning each telematics record a lane index $x_{i,t}$, we segment each lane longitudinally into fixed-length intervals, and each telematics record is assigned a longitudinal segment index $y_{i,t}$. The pair $(x_{i,t},y_{i,t})$ therefore identifies the unique cell visited by vehicle $i$ at time $t$, completing the cell-level representation $S_{i,t}=(x_{i,t},y_{i,t},v_{i,t})$. In our dataset, GPS timestamps arrive approximately every 3\,s. At typical freeway speeds (approximately 75\,mph or 33.5\,m/s), a vehicle travels about 100\,m within this interval, so we adopt 10\,m longitudinal segments when defining the index $y_{i,t}$. In practice, dividing each lane into roughly ten 10\,m cells provides sufficient spatial resolution without overcrowding consecutive samples in the same cell. Curvilinear sections of roadway are not problematic, as OSM provides dense centerline vertices that allow the curved alignment to be well approximated by multiple short polylines. Considering the inherent GPS noise, the resulting longitudinal segmentation remains robust under this geometric interpolation. This speed-adaptive rule yields uniformly occupied cells along each trajectory. By combining the lane index $x_{i,t}$ with the longitudinal segment index $y_{i,t}$, we obtain the final discretized state $S_{i,t}$ with minimal sampling bias.


\subsubsection{Vehicle Intention Distribution Extraction}

Through the discretization procedure described above, each raw telematics record 
$S^*_{i,t}$ is mapped to a cell-level state $S_{i,t}=(x_{i,t},y_{i,t},v_{i,t})$, 
providing a consistent spatiotemporal representation for modeling normal driving behavior.
This structure enables the extraction of a vehicle intention distribution, which captures 
the most likely next cell a vehicle will occupy under routine traffic conditions.

For each historical trajectory in $\mathcal{J}$ and for every time step $t$, 
we record the transition from the current cell $(x_{i,t},y_{i,t})$ to the next observed cell 
$(x_{i,t+1},y_{i,t+1})$ one timestamp later (3\,s). 
For a fixed origin cell $(x,y)$, all transitions of the form $(x,y)\rightarrow(x',y')$
are aggregated across all vehicles and all days in the 7-day historical window.  
Normalizing these aggregated counts yields an empirical transition probability:
\[
P\big((x,y)\rightarrow(x',y')\big),
\]
which represents the estimated statistical expectation that a vehicle in cell $(x,y)$ will move to cell $(x',y')$ in the next 3\,s interval.

The resulting vehicle intention distribution characterizes routine longitudinal and lateral movements across the corridor and forms the statistical baseline for our risk analysis, following a goal-prediction framework \citep{gan2025goal}.  
Transitions with high probability encode typical behavior, whereas any observed transition $S_{i,t}\to S_{i,t+1}$ that deviates substantially from these high-probability patterns is flagged as anomalous and treated as a potential risk factor.

\subsubsection{Threshold Determination}
In the offline stage, in addition to extracting the vehicle intention distribution from historical trajectories, we also use historical crash reports to calibrate the alert threshold for the real-time detection system. For each crash sample in the offline dataset, we run the complete risk-scoring procedure. This produces an accumulated risk map $R_t(x,y)$ for every cell $(x,y)$ over time. For each crash event, we record the maximum accumulated risk value attained within the affected spatiotemporal region. We then evaluate a range of candidate thresholds $T$ by comparing whether $R_t(x,y)\ge T$ would correctly identify the crash in a validation set. For each candidate threshold, we compute precision, recall, and the corresponding F1-score based on the binary detection outcomes. The optimal threshold $T^*$ is chosen as the value that maximizes the F1-score across all validation samples.

During real-time operation, the system continuously maintains the accumulated risk score $R_t(x,y)$ for every cell. An alert is triggered whenever any cell exceeds the offline-calibrated threshold $T^*$, providing a lane-level early-warning mechanism based on deviations from routine driving behavior.

\subsection{Online Detection Stage}
\label{sec:online}
The offline calibration stage provides two key components for real-time detection: 
the vehicle intention distribution $P((x,y)\!\to\!(x',y'))$, which represents normal driving behavior across the discretized grid $\mathcal{S}$, 
and the optimized crash alert threshold $T^*$. 
These components are then applied to incoming telematics data to dynamically compute a risk score. 
The risk scoring process translates observed trajectories 
$S^*_{i,t}=(\mathrm{lat}_{i,t},\mathrm{lon}_{i,t},v_{i,t})$ 
into discretized states 
$S_{i,t}=(x_{i,t},y_{i,t},v_{i,t})$ 
and evaluates deviations from normal driving patterns.  
At each time step $t$, a vehicle’s state at cell $(x_{i,t},y_{i,t})$ is evaluated by three complementary modules: 
transition anomaly detection, speed deviation analysis, and lateral maneuver risk assessment.  
Their weighted sum produces a comprehensive risk score capable of triggering real-time alerts.  

These module outputs are continuously accumulated on a spatial risk map 
$\{A_t(x,y):(x,y)\in\mathcal{S}\}$.  
To prevent these accumulated values from growing indefinitely, 
a reset indicator $\phi_t(x,y)$ is applied. 
When $\phi_t(x,y)=1$, the cell’s accumulated risk $A_t(x,y)$ is cleared, 
indicating that a new trajectory has passed through and the area is free of immediate hazards.  
This mechanism maintains numerical stability and ensures that the risk map reflects only recent, unrecovered anomalies.  
The following subsections describe each module in detail.

\subsubsection{Position Transition Risk}
To detect unlikely trajectory changes, we compare each cell transition of a vehicle 
from $S_{i,t}$ to 
$S_{i,t+1}$ 
against the vehicle intention distribution from offline analysis stage.  
Under normal conditions, vehicles follow high-probability transitions, whereas low-frequency transitions indicate abnormal movement. Therefore, transitions with low empirical frequency signal abnormal behavior.  We measure this surprise via a negative log‐likelihood transform because the negative log term penalizes the violation of highly probable transitions: 
when a transition with large expected probability \(P\) fails to occur, 
the term \(-\ln(1-P)\) produces a high risk value. :
\begin{align}
    R^{p}_{i,t}(x_{i,t},y_{i,t})
    &=
    -\ln\!\Bigl(1 - P\bigl((x_{i,t},y_{i,t})\!\to\!(x_{i,t+1},y_{i,t+1})\bigr)\Bigr).
\end{align}

Here, $P((x,y)\!\to\!(x',y'))$ is the empirical intention distribution estimated from historical trajectories.  
A high transition probability yields small risk, while rare transitions produce large penalties.  
A cutoff $\epsilon_p$ is applied to exclude transitions whose empirical probability is negligibly small. Such low probabilities may arise from data sparsity or estimation noise in the historical trajectories and typically correspond to infrequently
visited cells. Without filtering, these cells can accumulate small but persistent risk values over time during normal
driving. Applying this cutoff, together with the periodic cell-level reset described earlier, prevents these residual risks
from inflating the overall risk score and ensures numerical stability.

\subsubsection{Speed Deviation Risk}

Freeway crashes often involve sudden speed drops.  
For each discretized state $S_{i,t}=(x_{i,t},y_{i,t},v_{i,t})$, 
we flag scenarios where the instantaneous speed $v_{i,t}$ falls below a reference speed $v_{\mathrm{th}}$:

\begin{align}
R^{s}_{i,t}(x_{i,t},y_{i,t}) &=
\begin{cases}
\dfrac{v_{\mathrm{th}} - v_{i,t}}{v_{\mathrm{th}}}, & v_{i,t} \le v_{\mathrm{th}},\\[6pt]
0, & v_{i,t} > v_{\mathrm{th}}.
\end{cases}
\end{align}

\subsubsection{Lateral Maneuver Risk}
Frequent lane changes within a confined area can signal lane blockage caused by a crash or hazard. 
When a lane is temporarily occupied, vehicles often move laterally to adjacent lanes to bypass the obstruction and then return once the lane becomes clear. 
Hence, consecutive lane-change events detected at the same location over a short period indicate potential incident-related disturbances. 
We detect a lane-change event by comparing successive lane indices $x_{i,t} \neq x_{i,t-1}.$ and assign a risk score for each occurrence:
\begin{align}
R^{\ell}_{i,t}(x_{i,t},y_{i,t}) 
= \mathbf{1}\{x_{i,t} \neq x_{i,t-1}\}.
\end{align}
The indicator $\mathbf{1}\{\cdot\}$ equals one when a vehicle moves laterally between lanes in one timestep.  
Cells where lane-change events occur repeatedly over consecutive timesteps accumulate higher risk, 
reflecting the likelihood of temporary blockage or crash-induced detours in that segment.

\subsubsection{Total Risk Score}

The risk components defined above represent vehicle-level quantities evaluated at each time step for the discretized state 
$S_{i,t}$.  
To obtain a spatially aggregated measure over the discretized grid, 
each vehicle’s instantaneous risk is mapped to its corresponding spatial cell 
$(x_{i,t},y_{i,t})\in\mathcal{S}$ 
and summed with those of all vehicles currently occupying that same cell.  
This aggregation produces a cell-level representation of abnormality that reflects collective traffic behavior:

\begin{align}
\bar{R}^{m}_{t}(x,y) 
= \sum_{i:\,x_{i,t}=x,\,y_{i,t}=y} 
  R^{m}_{i,t}(x_{i,t},y_{i,t}),
\qquad m\in\{p,s,\ell\}.
\end{align}

Here, $m$ indexes the three modules: position transition risk, speed deviation risk, 
and lateral maneuver risk.  
Cells with many vehicles exhibiting similar anomalies naturally accumulate larger cell-level risks, 
providing robustness against noise in individual trajectories. These aggregated cell-level risks are then combined to form the total instantaneous risk for each cell:
\begin{align}
\bar{R}_{t}(x,y)
&= 
w_p\,\bar{R}^{p}_{t}(x,y)
+ w_s\,\bar{R}^{s}_{t}(x,y)
+ w_{\ell}\,\bar{R}^{\ell}_{t}(x,y),
\end{align}
where the nonnegative coefficients $(w_p,w_s,w_{\ell})$ are calibrated offline.  

To capture abnormal behavior that persists over time, we define the accumulated risk for each cell $(x,y)$ by
$A_t(x,y)$, initialized with $A_0(x,y)=0$.  
The accumulated risk grows as long as anomalies continue to occur near that location.  
However, when a new trajectory passes through a cell, that cell is directly observed to be navigable, and thus temporarily hazard-free. To reflect this, a cell-level reset mechanism is introduced:

\begin{align}
A_t(x,y)
=
\bigl(1-\phi_t(x,y)\bigr)\,
\bigl[A_{t-1}(x,y)+\bar{R}_{t}(x,y)\bigr],
\end{align}

where $\phi_t(x,y)\in\{0,1\}$ is a reset indicator equal to 1 iff there exists some vehicle $i$ such that 
$(x_{i,t},y_{i,t})=(x,y)$ at time $t$.  
When $\phi_t(x,y)=1$, the cell risk is cleared to zero because the region has been directly traversed and verified as safe.  
This prevents false escalation of long-term accumulated risk and maintains numerical stability.

A lane-level crash alert is issued the first time
\begin{align}
A_t(x,y) \ge T^*,
\end{align}
where $T^*$ is the crash threshold selected during offline calibration using historical crash reports.  
This additive and accumulative structure enables the system to fuse short-term anomalies and sustained abnormal behavior into a coherent spatiotemporal risk metric.  
Whenever the accumulated risk in any cell exceeds $T^*$, the system immediately emits a crash alert and reports the cell $(x,y)$ as the detected crash location. We summarized the whole methodology into the following algorithm \ref{alg:crash_detection}.

\begin{algorithm}[tbp]
\caption{Crash Detection: Offline and Online Stages}
\label{alg:crash_detection}
\begin{algorithmic}[1]
\Require historical telematics dataset $\mathcal{J} = \{S^*_{i,t}\}$, 
historical crash records, lane width $w$, lane count $n$, 
transition cutoff $\epsilon_p$, speed baseline $v_{\mathrm{th}}(x,y)$, 
module weights $(w_p, w_s, w_{\ell})$.

\Statex
\Statex \textbf{Offline Stage}
\State Discretize all records 
    $S^*_{i,t} = (\text{lat}_{i,t}, \text{lon}_{i,t}, v_{i,t})$
    into 
    $S_{i,t} = (x_{i,t}, y_{i,t}, v_{i,t})$
    via map-matching and longitudinal segmentation.
\State Estimate empirical transition probabilities
    $P\big((x,y)\!\to\!(x',y')\big)$ 
    from non-crash portions of $\mathcal{J}$.
\State Estimate normal speed baselines $v_{\mathrm{th}}(x,y)$ for each cell.
\State Determine alert threshold $T^*$ by maximizing validation F1-score based on official crash reports.
\State Calibrate risk module weights $(w_p, w_s, w_{\ell})$.

\Statex
\Statex \textbf{Online Stage}
\State Initialize accumulated risk $R_0(x,y) \gets 0$ for all $(x,y)\in\mathcal{C}$.
\For{each time step $t = 1, 2, \dots$}
  \State Receive telematics records 
        $\{S^*_{i,t} = (\text{lat}_{i,t}, \text{lon}_{i,t}, v_{i,t}) : i \in \mathcal{I}_t\}$.
  \State Discretize each record to 
        $S_{i,t} = (x_{i,t}, y_{i,t}, v_{i,t})$ 
        and retrieve $(x_{i,t-1}, y_{i,t-1})$ if available.

  \Statex \quad \emph{(Per-vehicle instantaneous risks at $(x_{i,t},y_{i,t})$)}
  \For{each vehicle $i \in \mathcal{I}_t$}
    \State \textbf{Position-transition risk:}
    \[
      R^{p}_{i,t}(x_{i,t},y_{i,t}) =
      \begin{cases}
        -\ln \!\Bigl(
          1 - P\bigl((x_{i,t-1},y_{i,t-1}) \to (x_{i,t},y_{i,t})\bigr)
        \Bigr), 
        & \text{if } P > \epsilon_p,
        \\[4pt]
        0, & \text{otherwise}.
      \end{cases}
    \]

    \State \textbf{Speed risk:}
    \[
      R^{s}_{i,t}(x_{i,t},y_{i,t}) =
      \max\!\Bigl(
        0,\;
        \frac{v_{\mathrm{th}}(x_{i,t},y_{i,t}) - v_{i,t}}
             {v_{\mathrm{th}}(x_{i,t},y_{i,t})}
      \Bigr).
    \]

    \State \textbf{Lateral maneuver risk:}
    \[
      R^{\ell}_{i,t}(x_{i,t},y_{i,t})
      = \mathbf{1}\{\, x_{i,t} \neq x_{i,t-1} \,\}.
    \]
  \EndFor

  \Statex \quad \emph{(Aggregate per-vehicle risks to cell level)}
  \For{each occupied cell $(x,y)$ at time $t$}
     \State $\bar{R}^{m}_t(x,y) \gets 
       \displaystyle\sum_{i:\,x_{i,t}=x,\,y_{i,t}=y}
       R^{m}_{i,t}(x,y)$ 
       \quad for $m\in\{p,s,\ell\}$
     \State $\bar{R}_t(x,y) \gets 
        w_p\,\bar{R}^{p}_t(x,y) 
      + w_s\,\bar{R}^{s}_t(x,y) 
      + w_{\ell}\,\bar{R}^{\ell}_t(x,y)$
  \EndFor

  \Statex \quad \emph{(Update accumulated risk map and detect crashes)}
  \For{each cell $(x,y)\in\mathcal{C}$}
     \State $\phi_t(x,y) \gets \mathbf{1}\{\exists\,i:\,x_{i,t}=x,\; y_{i,t}=y\}$
        \Comment{reset if any vehicle enters $(x,y)$}
     \State $R_t(x,y) \gets 
        (1-\phi_t(x,y))\,[\, R_{t-1}(x,y) + \bar{R}_t(x,y) \,]$
     \If{$R_t(x,y)\ge T^*$}
       \State \Call{EmitCrashAlert}{$(x,y),\, t$}
     \EndIf
  \EndFor
\EndFor

\end{algorithmic}
\end{algorithm}

\section{Experiment}
\label{SEC_EXPERIMENT}
\subsection{Dataset and Experimental Setup}
The dataset used in this study spans the entire United States and covers two 
meteorologically distinct, continuous one-week periods: a summer week 
(August 4--10, 2024) and a winter week (January 9--15, 2024). 
The nationwide summer dataset alone contains approximately 11.4 terabytes of 
telematics data stored across 208{,}840 Parquet files, totaling about 
$1.1\times10^{11}$ GPS observations sampled at 3-second intervals. 
To complement this large-scale data, the winter dataset adds an 
additional 215 gigabytes of high-resolution telematics records covering the 
entire state of Wisconsin. 
This combination of nationwide coverage, fine-grained temporal sampling, and 
cross-seasonal variation provides a robust basis for evaluating real-time 
crash detection. The data span a wide range of roadway geometries, regional 
driving behaviors and traffic densities, allowing rigorous 
assessment of the system's generalization and operational robustness.

\begin{figure}[t]
  \centering
  \includegraphics[width=0.75\textwidth]{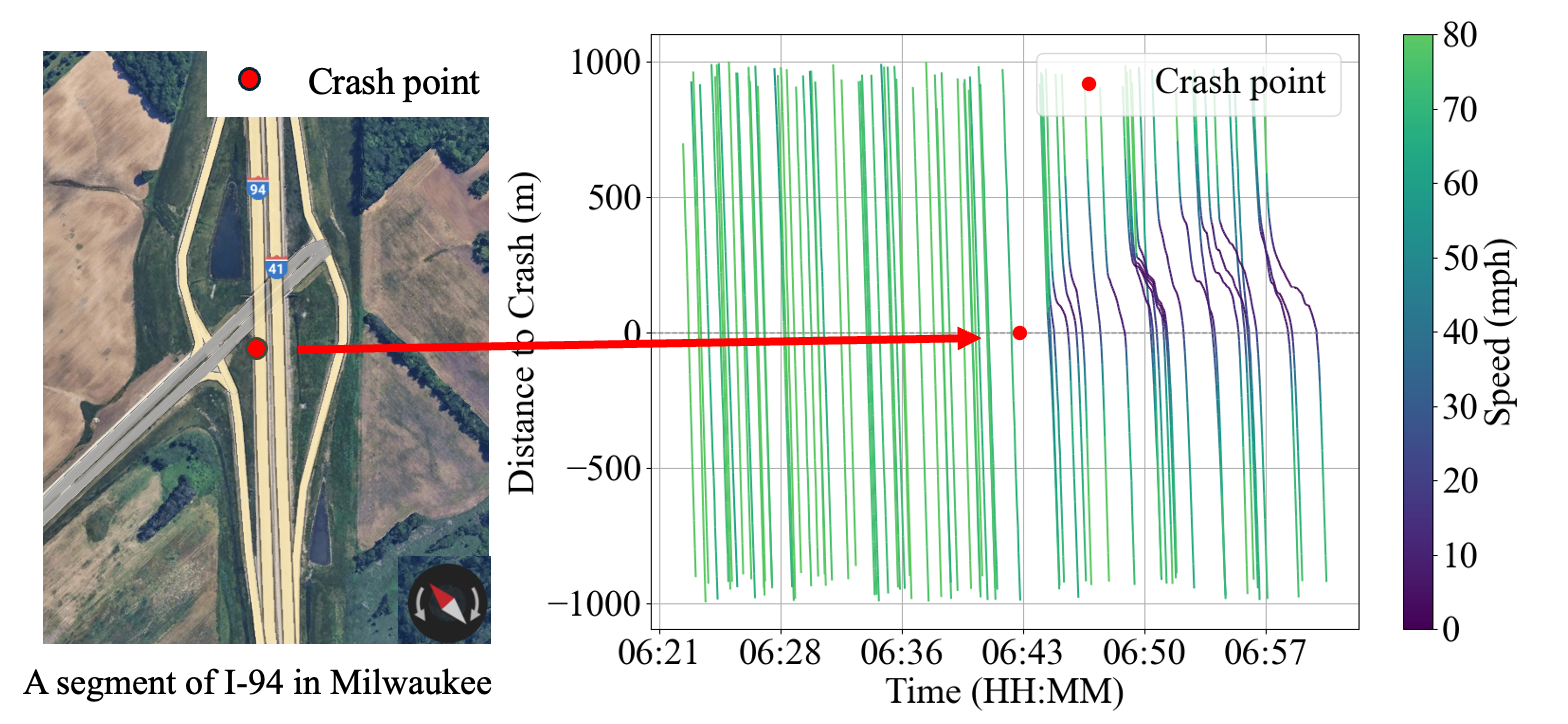}
     \caption{Crash location and corresponding time-space diagram}
  \label{fig:crash_distribution}
\end{figure}

\begin{figure}[htbp]
  \centering
  \includegraphics[width=\textwidth]{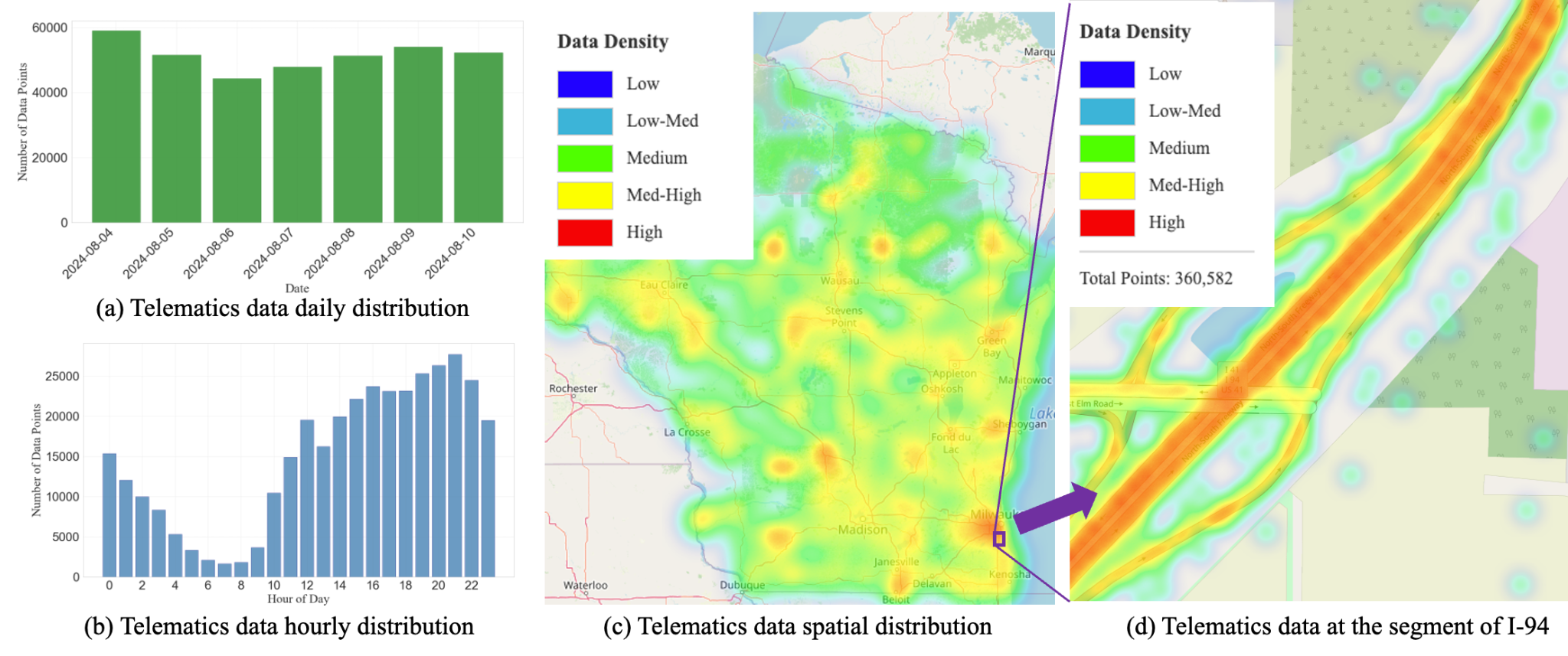}
  \caption{Telematics data coverage and density for the freeway segment}
  \label{fig:tele_data_distribution}
\end{figure}

For this study, we restrict our analysis to vehicle trajectories located within Wisconsin, as all crash reports originate from this region.
To avoid confounding delays caused by signalized intersections, we further restrict our attention to crashes on freeway segments. Figure Figure \ref{fig:crash_distribution} illustrates a representative crash case in Milwaukee during August 2024. The left figure shows the spatial location of the crash along the I-94 freeway segment. The right figure presents the corresponding time–space trajectories of vehicles approaching the crash site, with the crash occurring at 06:43 on August 5, 2024. Vehicle trajectories are extracted within a fixed radius (approximately 200m) centered at the crash point, covering a 25-minute window before and after the incident. Each trajectory is color-coded by instantaneous speed. The evident slowdown and clustering of trajectories downstream clearly indicate the congestion induced by the crash.
Based on the filtered trajectories obtained around the crash point, we further construct the vehicle intention distribution for this segment. The telematics data are exceptionally rich. Over the seven‐day window, more than 360 000 trajectory points were recorded, with the majority concentrated in the main travel lanes (see Figure \ref{fig:tele_data_distribution}(d)). Figure \ref{fig:tele_data_distribution}(a) presents the daily distribution of GPS samples, and Figure \ref{fig:tele_data_distribution}(b) shows the hourly distribution on the crash day (August 5th). In addition, Figure \ref{fig:tele_data_distribution}(c) visualizes the statewide spatial density map of telematics points, highlighting high-coverage regions along major freeway corridors. Together, these distributions demonstrate the dataset’s comprehensive temporal and spatial coverage necessary for robust crash detection.


\subsection{Discretization and vehicle intention distribution Extraction}
\begin{figure}[htbp]
  \centering
  \includegraphics[width=\textwidth]{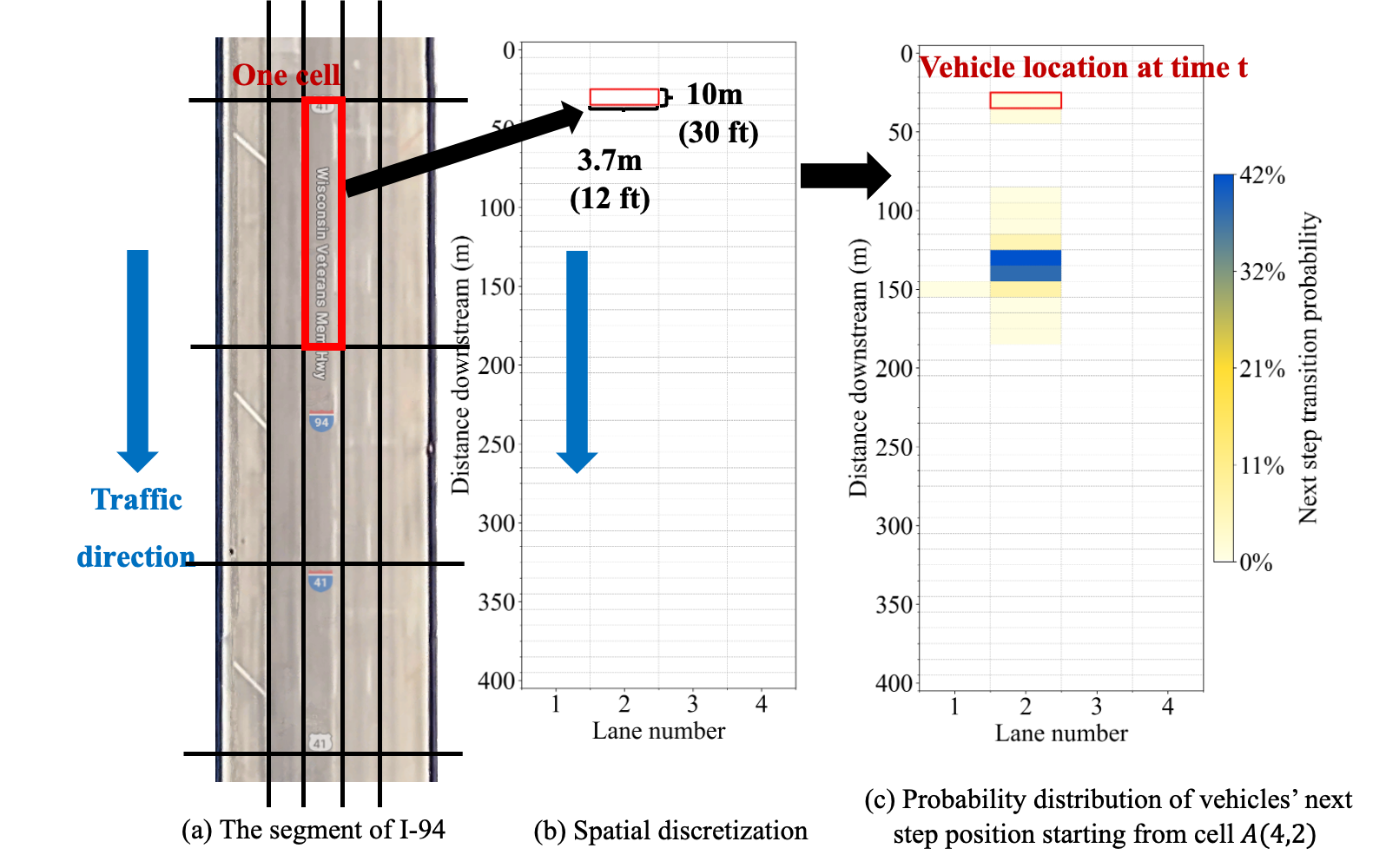}
  \caption{Illustration of spatial–temporal discretization and vehicle intention distribution extraction.}
  \label{fig:discretization_and_behavior_model}
\end{figure}

In the offline stage, following Section~\ref{sec:offline} and Algorithm~\ref{alg:lane_matching}, we first discretize the area surrounding the crash location as shown in Figure~\ref{fig:discretization_and_behavior_model}(a) and (b). In the spatial dimension we discretize laterally by lane and longitudinally using speed-dependent segments. For lateral segmentation, if OSM provides explicit lane-width tags, we use those values; otherwise we default to the U.S. freeway standard lane width of 3.7 m~\citep{hancock2013policy}. For longitudinal segmentation in this example, with a posted speed limit of 75 mph ($\approx$ 33.5 m/s), we use 10 m cells so that a vehicle travels roughly 10 cells between three-second samples. This rule-based design balances spatial resolution against the sparsity of three-second telemetry.  Considering the noise of telematics data, moderate variations in lane width or segment length have minimal effect on detection performance.

Next, we apply this process to the 360 000+ trajectory points collected over the seven-day period. After filtering by heading north degree from telematics data and snapping to the derived road centerline based on OSM road geometry, we isolate the crash-affected segment freeway. Using the vehicle intention distribution extraction method in Section~\ref{sec:offline}, we count transitions from each source cell and normalize them to derive the vehicle intention distribution. Figure~\ref{fig:discretization_and_behavior_model}(c) highlights one source cell (red outline). The resulting spatial distribution shows the likelihood of transitions after three seconds, with darker blue indicating higher probability. Every cell in the discretized grid is equipped with this vehicle intention distribution.  

As part of the offline stage, we replay the algorithm \ref{alg:crash_detection} using telematics trajectories and official crash reports from the summer dataset (August 2024) to calibrate the alert threshold. Specifically, we sweep~\(T\) across the range of observed accumulated risk scores and, for each candidate, compute precision, recall, and F1‐score. Table~\ref{tab:f1_scores_full} summarizes these results, showing that the maximum F1‐score of~0.857 occurs at~\(T = 30\), which is adopted as the final alert threshold~\(T^*\). 

\begin{table}[ht]
\centering
\caption{Threshold vs.\ Precision, Recall and F$_1$‐Score}
\label{tab:f1_scores_full}
\begin{tabular}{rrrr}
\toprule
Threshold & Precision & Recall & F$_1$‐score \\
\midrule
0.000   & 0.186 & 1.000 & 0.313 \\
16.084  & 0.619 & 1.000 & 0.765 \\
18.076  & 0.600 & 0.923 & 0.727 \\
20.442  & 0.632 & 0.923 & 0.750 \\
23.142  & 0.667 & 0.923 & 0.774 \\
26.410  & 0.706 & 0.923 & 0.800 \\
28.182  & 0.750 & 0.923 & 0.828 \\
\textbf{30.000}  & \textbf{0.800} & \textbf{0.923} & \textbf{0.857} \\
39.008  & 0.750 & 0.692 & 0.720 \\
60.000  & 0.727 & 0.615 & 0.667 \\
62.300  & 0.889 & 0.615 & 0.727 \\
90.000  & 0.875 & 0.538 & 0.667 \\
120.000 & 1.000 & 0.538 & 0.700 \\
150.000 & 1.000 & 0.462 & 0.632 \\
\bottomrule
\end{tabular}
\end{table}

\subsection{Online Risk Score Dynamics}

\begin{figure}[htbp]
  \centering
  \includegraphics[width=\textwidth]{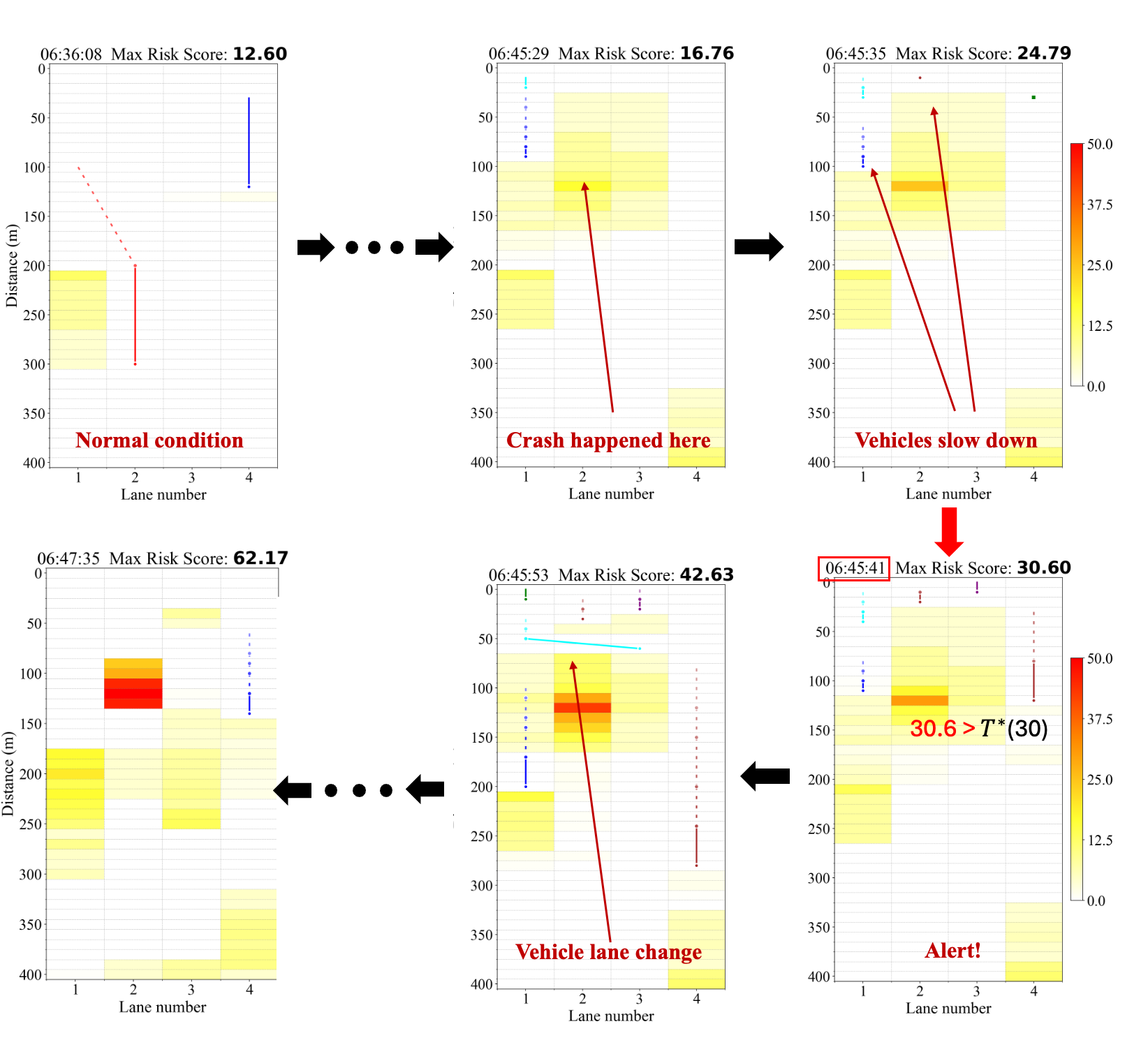}
  \caption{Sequential lane-level risk map figuress with annotated maximum risk scores.}
  \label{fig:risk_map_revolution}
\end{figure}


In the online stage, each incoming telematics record 
$S^*_{i,t}=(\mathrm{lat}_{i,t},\mathrm{lon}_{i,t},v_{i,t})$ 
is immediately converted into its discretized state 
$S_{i,t}=(x_{i,t},y_{i,t},v_{i,t})$ 
using the lane-matching algorithm (Algorithm~\ref{alg:lane_matching}) 
together with longitudinal segmentation.  
Once the spatial cell $(x_{i,t},y_{i,t})$ is obtained, three complementary risk components 
$R^{p}_{i,t}$, $R^{s}_{i,t}$, and $R^{\ell}_{i,t}$ 
are computed at that cell.  
Because these components contribute differently to crash inference, 
we weight them with factors $w_p=3$, $w_s=2$, and $w_{\ell}=4$, respectively, 
reflecting the relative importance of abrupt lateral maneuvers (highest weight), 
unusual position transitions, and abnormal speed deviation.

The instantaneous risk contributed by vehicle $i$ at time $t$ is therefore
\[
  R_{i,t}(x_{i,t},y_{i,t}) 
  = 
  w_p\,R^{p}_{i,t}(x_{i,t},y_{i,t})
  + w_s\,R^{s}_{i,t}(x_{i,t},y_{i,t})
  + w_{\ell}\,R^{\ell}_{i,t}(x_{i,t},y_{i,t}).
\]

These per-vehicle risks are aggregated across all vehicles occupying the same cell 
to obtain the cell-level risk $\bar{R}_t(x,y)$, 
which is then accumulated through time via the risk map $A_t(x,y)$.  
If a vehicle traverses cell $(x,y)$ at time $t$, 
the reset indicator $\phi_t(x,y)=1$ clears the accumulated risk, 
indicating that the local hazard has been resolved.  
Whenever any cell satisfies $A_t(x,y)\ge T^*$, 
the system immediately triggers a lane-level crash alert.

Figure~\ref{fig:risk_map_revolution} illustrates the evolution of the lane-level risk map for the crash event discussed in the online stage. The first frame in this figure (06:36:08) shows normal traffic conditions with uniformly low risk levels.

In the sequence of four frames of Figure~\ref{fig:risk_map_revolution} (06:45:29 to 06:45:53), the system begins to capture the disturbance once new telematics records become available. Because only a very small number of vehicles passed the crash location between 06:43 and 06:45 due to data sparsity, the system could not observe any abnormal behavior during that window. When the density of observations increases at 06:45:29, likely due to the onset of congestion, the first visible signs of abnormal behavior appear, including simultaneous slowdowns and lane changes that are inconsistent with the offline intention distribution. These anomalies cause the risk to rise quickly, and by 06:45:41 the maximum score reaches \(30.6 > T^*{=}30\), which triggers an alert approximately 1.2 minutes before the official report time.

The final frame of Figure~\ref{fig:risk_map_revolution} (06:47:35) shows continued growth of the risk around the crash location, matching the evolution of the actual crash and refining the detected lane-level location over time.

\subsection{Overall Performance}

\begin{figure}[htbp]
  \centering
  \includegraphics[width=\textwidth]{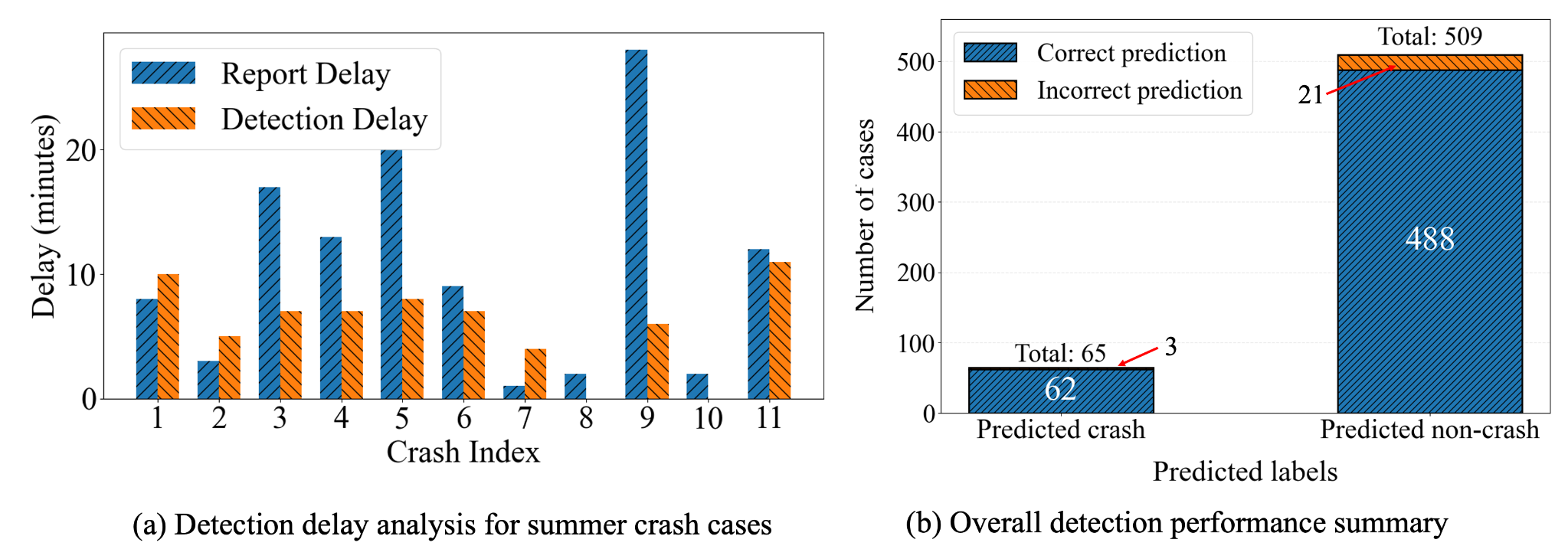}
  \caption{Detection performance evaluation.}
\label{fig:detection_results}
\end{figure}

To further evaluate the system performance, we extend the online validation to include both the summer and winter datasets. The alert threshold \(T^*\) and component weights \((w_p, w_l, w_s)\), calibrated from the summer data, are applied directly to the winter dataset without retraining or parameter adjustment. As shown in Fig.~\ref{fig:detection_results}(a), a portion of the crash cases can be detected earlier than the official report time, demonstrating that the system is capable of identifying emerging disturbances in advance when sufficient telematics observations are available.
Table~\ref{tab:confusion_matrix}
summarizes the combined detection results across all evaluated cases. Out of 83 crash events, 62 are successfully detected, yielding a crash detection rate of 74.7\%. Among 491 non-crash cases, only three false alarms occur, resulting in a precision of 95.4\% and an overall F1 score of 0.84. The results are also shown in Fig.~\ref{fig:detection_results}(b). 
\begin{table}[t]
\centering
\caption{Confusion matrix of crash detection results}
\label{tab:confusion_matrix}
\begin{tabular}{cc|cc|c}
\toprule
\multicolumn{2}{c|}{\textbf{}} & \multicolumn{2}{c|}{\textbf{Actual}} & \textbf{Total} \\
\cline{3-4}
\multicolumn{2}{c|}{\textbf{Predicted}} 
& Crash & Non-crash & \\
\midrule
Crash     &  & 62  & 3   & 65 \\
Non-crash &  & 21  & 488 & 509 \\
\midrule
Total     &  & 83  & 491 & 574 \\
\bottomrule
\end{tabular}
\end{table}
It is worth noting that the missed crashes mainly correspond to minor incidents that produce no observable disturbance in lane-level trajectories, such as vehicles pulling over, mechanical failures, or low-speed fender-benders occurring on the shoulder. For example, one undetected case in the crash reports describes a vehicle pulling onto the right shoulder and catching fire without affecting mainline traffic flow. Such events, although officially classified as crashes, do not materially influence lane-level behavior and therefore fall outside the scope of real-time traffic impact detection. Consequently, the proposed framework remains highly effective for safety-critical crashes that generate measurable be

\section{Conclusion}
\label{SEC_CONCLUSION}
This study introduces a novel real-time, lane-level crash detection framework that relies exclusively on sparse telematics data and road geometry from OSM. In the offline stage, we perform low-cost lane matching using a vector cross-product method and discretize vehicle trajectories into spatial cells to mitigate data sparsity. During the online stage, each incoming telematics record is evaluated for position transition anomalies, speed deviations, and abnormal lane-change maneuvers, and these risk components are accumulated in a lane-specific map. When the accumulated risk in any cell surpasses the threshold determined offline via F1‐score optimization, the system issues an immediate alert with precise lane-level localization, enabling timely warnings that can help prevent secondary collisions and reduces associated economic losses and casualties.

To comprehensively validate the framework, we conducted experiments on combined summer (August 2024) and winter (January 2024) datasets covering freeway segments across Wisconsin. A total of 83 crash events and 491 non-crash control periods were analyzed. The proposed method successfully detected 62 crashes while producing only three false alarms, yielding a detection rate of 74.7\%, a precision of 95.4\%, an overall F1‐score of 0.84, and an accuracy of 96.0\%. Notably, 13\% of the detected crashes were identified 3 minutes before the official report time. The missed cases were primarily minor incidents that caused no measurable disturbance in lane-level trajectories, whereas all major lane-blocking crashes were accurately localized to the correct lanes. These results demonstrate that the framework generalizes well across seasons and maintains robust detection and localization performance under diverse traffic and environmental conditions.


Future work will explore learning-based intention models that integrate kinematic structure and historical trajectory features. At present, parameters such as risk module weights are determined through empirical tuning, and learning-based methods could automatically optimize these quantities from data. As the dataset grows, threshold calibration may also be refined to account for peak and off-peak periods, roadway types, and regional driving patterns. With greater computational resources, the framework could further adapt its parameters in real time to improve detection speed and accuracy.

\section{Acknowledgments }
We gratefully acknowledge the support and guidance provided by the Center for Connected and Automated Transportation project “Evaluation of Vehicle Telematics and Infrastructure-based Connected Vehicle Data for Real-Time Safety and Mobility Applicationin" in this work. 

\bibliographystyle{cas-model2-names}

\bibliography{cas-refs}

\end{document}